\documentclass[aps,prl,reprint,superscriptaddress]{revtex4-1}

\usepackage{amsmath,bm,mathrsfs,color}
\usepackage{graphicx}

\newcommand{\scr}[1]{\mathscr{#1}}
\newcommand{\br}{\bm{r}}
\newcommand{\bp}{\bm{p}}

\newcommand{\bt}{\bm{t}}
\newcommand{\bu}{\bm{u}}

\newcommand{\kp}{\bm{k}\cdot \bm{p}}
\newcommand{\D}{\Delta}
\newcommand{\up}{\uparrow}
\newcommand{\down}{\downarrow}
\newcommand{\wc}{\omega_{\rm c}}

\begin{document}


\title{Origin of the large anisotropic g-factor of holes in bismuth}


\author{Yuki Fuseya}
\email[]{fuseya@pc.uec.ac.jp}
\affiliation{Department of Engineering Science, University of Electro-Communications, Chofu, Tokyo 182-8585, Japan}

\author{Zengwei Zhu}
\author{Beno\^{\i}t Fauqu\'e}
\affiliation{
LPEM (UPMC-CNRS), Ecole Su\'perieure de Physique et de Chimie Industrielles, 75005 Paris, France
}

\author{Woun Kang}
\affiliation{
Department of Physics, Ewha Womans University, Seoul 120-750, Korea
}

\author{Bertrand Lenoir}
\affiliation{
Institut Jean Lamour (UMR 7198 CNRS, Nancy Universit\'e, UPVM), Ecole Nationale Sup\'erieure des Mines de Nancy, 54042 Nancy, France
}

\author{Kamran Behnia}
\affiliation{
LPEM (UPMC-CNRS), Ecole Su\'perieure de Physique et de Chimie Industrielles, 75005 Paris, France
}

\date{\today}

\begin{abstract}
The  ratio of the Zeeman splitting to the cyclotron energy ($M=\D E_{\rm Z}/\hbar \omega_{\rm c}$) for hole-like carriers in bismuth has been quantified with a great precision by many experiments performed during the past five decades. It exceeds 2 when the magnetic field is along the trigonal axis and vanishes in the perpendicular configuration. Theoretically, however,  $M$ is expected to be isotropic and equal to unity in a two-band Dirac model. We argue that a solution to this half-a-century-old puzzle can be found by extending the $\kp$ theory to multiple bands. Our model not only gives a quantitative account of magnitude and anisotropy of $M$  for hole-like carriers in bismuth, but also explains its contrasting evolution with antimony doping pressure, both probed by new experiments reported here. The present results have important implications for the magnitude and anisotropy of $M$ in other systems with strong spin-orbit coupling.
\end{abstract}


\maketitle


	Spin-orbit interaction (SOI) is a central issue in contemporary solid state physics. It is an automatic consequence of the Dirac theory, and  well understood for a lonely electron in presence of a single atomic potential. In crystals, however, a diversity of SOI effects arise due to variety of the crystalline potentials. The effect also strongly depends on the momentum since the magnitude of SOI is set by $(\hbar/4m^2 c^2)\bm{\sigma}\cdot \bm{\nabla}V(\br)\times \bp$. (Here $V(\br)$ is the crystalline potential, $\bp$ is a momentum, and $\bm{\sigma}$ is the Pauli spin matrix.) It is thus basically difficult to study the effect of SOI for various materials and various $k$-points of the Brillouin zone by a universal approach.
	
	One signature of crystalline SOI is its impact on the ratio of the Zeeman splitting $\D E_{\rm Z}$ to the cyclotron energy $\hbar \wc$, dubbed $ M \equiv \D E_{\rm Z}/\hbar \wc$. The crystalline SOI appears as an antisymmetric part with respect to commutation of momentum operators in the Hamiltonian under a magnetic field \cite{Luttinger1956,Roth1959,Cohen1960,Yafet1963,Wolff1964,Fuseya2015}. Its eigenvalue corresponds to the effective Zeeman energy with an anisotropic effective g-factor $\tilde{g}$, while the eigenvalue of the symmetric part corresponds to the cyclotron energy with an anisotropic effective mass. Therefore, the impact of the crystalline SOI can be characterized by the relative energy scale of the crystalline SOI to the kinetic energy, i.e., $M=\D E_{\rm Z}/\hbar \wc$. This ratio can be accurately determined by experiment in those cases in which quantum oscillations simultaneously detect  successive Landau levels as well as  same-index sub-levels with opposite spins \cite{Smith1964,Edelman1976,Bompadre2001,Behnia2007_PRL,ZZhu2011,ZZhu2012,Kohler1975,Fauque2013}. The energy levels for different cases and their corresponding $M$ are illustrated in Fig. \ref{Fig1A}, and the experimental values of $M$ in three different systems are listed in Table \ref{t1}. As seen in this Table, when SOI is weak (i.e., in the case of graphite), $M$ is much smaller than unity. One can show that $M$ is exactly equal to unity (and so never exceeds unity) for any direction of magnetic field when a large SOI strongly couples two bands based on the two-band model, which is equivalent to the Dirac Hamiltonian \cite{Cohen1960,Wolff1964,Fuseya2015}. This agrees quite well with the experiments on the $L$-point of bismuth.
	
\begin{figure}
\begin{center}
\includegraphics[width=7cm]{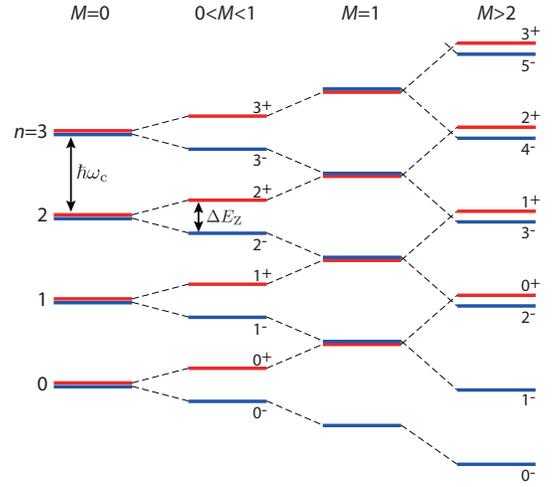}
\end{center}
\label{Fig1A}
\caption{\label{Fig1A}Energy levels under a magnetic field for different ratio of the Zeeman splitting to the cyclotron energy, $M =\D E_{\rm Z}/\hbar \wc$. }
\end{figure}

\begin{table*}
 \caption{\label{t1}Ratio of the Zeeman splitting to the cyclotron energy for some compounds. The cyclotron mass \cite{Soule1964,Roth1959,Vdovin1986,ZZhu2011}, effective g-factor \cite{Schneider2010,Roth1959,Vdovin1986,ZZhu2011}, and atomic SOI \cite{Yafet1963} are also listed. For graphite, only values with the magnetic field parallel to the $c$-axis are listed. The range of values for Bi expresses their anisotropy.}
 \begin{ruledtabular}
 \begin{tabular}{cccccc}
	 & \multicolumn{2}{c}{Graphite} & InSb & \multicolumn{2}{c}{Bi}\\ \hline
position & $K$ (ele.) & $H$ (hole) & $\Gamma$ & $L$ (ele.) & $T$ (hole)\\
$m_{\rm c} /m $&  0.038 & 0.057 & 0.014 & 0.0019-0.027 & 0.068-0.22 \\
$\tilde{g}$ & 2.5 & 2.5 & 52 & 74-1060 & 0.79-63 \\
$M=m_{\rm c} \tilde{g} /2m$& 0.048& 0.073 & 0.36 & 0.9-1.0 & 0.0-2.12 \\
atomic SO (eV) & 0.005 & 0.005  & 0.27, 0.68 & 1.8 & 1.8 
 \end{tabular}
 \end{ruledtabular}
 \end{table*}
 
	However, in the case of holes at $T$-point, there are two puzzling features: (i) $M$ is extremely anisotropic ($\simeq 0$ for $B\perp$ trigonal axis); (ii) It largely exceeds unity in one configuration ($ =2.12$ for $B\parallel$ trigonal axis) as shown later. These two puzzles emerged from numerous experiments starting half a century ago [see supplementary materials (SM)] in absence of any satisfactory explanations \cite{Smith1964,Edelman1976,Bompadre2001,Behnia2007_PRL,ZZhu2011,ZZhu2012}. In other words, the two-band approach completely fails to give its adequate value.

	In this Letter, we show that a satisfactory solution to these longstanding puzzles can be found by going beyond the two-band Dirac model. Furthermore, we present new experimental results on the evolution of $M$ for holes in bismuth with pressure and antimony doping. Rather counterintuitively, these two alternative ways to reduce carrier concentration shift the magnitude of $M$ in opposite directions. We show this can be quantitatively explained by the present approach.

\begin{figure}
\begin{center}
\includegraphics[width=7cm]{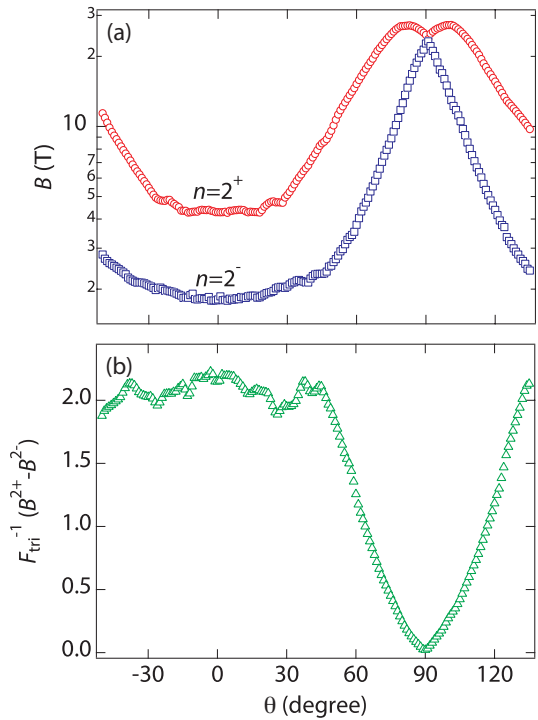}
\end{center}
\caption{\label{Fig2}Angular dependence of a Kramers doublet in bismuth. (a) Magnetic field at which $n=2^+, 2^-$ Landau levels are evacuated as a function of the orientation of the magnetic field. $\theta$ is the angle between the magnetic field and the trigonal axis. Symbols represent the field at which peaks are resolved in the Nernst quantum oscillation data (Ref.\cite{ZZhu2012}). (b) Angular dependence of  $F_{\rm tri}^{-1} (B^{2+}-B^{2-})$ from panel (a).
}
\end{figure}

	Numerous experiments have quantified the magnitude of the g-factor (and therefore $M$) for both electrons and holes in bismuth \cite{Smith1964,Edelman1976,Bompadre2001,Behnia2007_PRL,ZZhu2011,ZZhu2012}. The large atomic SOI of bismuth ($\sim 1.8$ eV) overwhelmingly dominates all other energy scales including the Fermi energy ($\simeq 28$ meV) and the band gap ($\simeq 15$ meV). Therefore, the crystalline SOI dramatically affects the electronic structure and this leads to a very complex hierarchy between $\hbar \wc$ and $\D E_{\rm Z}$ for different carriers and different orientations of magnetic field. In the case of electrons at the $L$-point of the Brillouin zone, $M \simeq 1$ with little dependence on the field orientation. In other words, in spite of the extreme anisotropy of both the cyclotron mass $m_{\rm c}$ and $\tilde{g}$ \cite{ZZhu2011}, $\hbar \wc$ and $\D E_{\rm Z}$ remain almost equal to each other. This property of the $L$ electrons is quite well-understood based on the two-band model \cite{Cohen1960,Wolff1964,Fuseya2015} and were employed to give a quantitative account of the complex Landau spectrum of $L$ electrons \cite{ZZhu2011,ZZhu2012}. The puzzling features experimentally observed on holes at the $T$-point are shown in Fig. \ref{Fig2} (a), which shows the angular dependence of two Nernst peaks corresponding to energy levels $n=2^{\pm}$, where $\pm$ indicates the degree of freedom of the Kramers doublet \cite{ZZhu2012}. The angular dependence of $M$ can be deduced by plotting $F_{\rm tri}^{-1}(B^{2+}-B^{2-})$, which is shown in Fig. \ref{Fig2} (b), where $B^{2\pm}$ is the magnetic field of the Nernst peak for $n=2^{\pm}$ and $F_{\rm tri}$ is the oscillation frequency for $B\parallel $ trigonal axis. For $B\parallel $ trigonal, this value is exactly the same as $M$, and so we obtain $M=2.12$ (cf. Fig. \ref{Fig1A}). For $B\perp$ trigonal, on the other hand, $F_{\rm tri}^{-1}(B^{2+}-B^{2-})$ becomes almost zero indicating $M\simeq0$.

	Next we give a general theory on the ratio $M$. We start from one-electron Hamiltonian in presence of strong SOI and apply $\kp$ theory to a multiband ($n$-band) system under a magnetic field taking into account the SOI in a fully relativistic (non-perturbative) way. The obtained Hamiltonian is written in terms of $2n \times 2n$ matrix for an $n$-band system. (See SM for details.) In order to study the electromagnetic properties of electrons belonging to a particular band, we decouple $2\times 2$ Hamiltonian, $\scr{H}_{n=0}$, from the other bands by using the L\"owdin's unitary transformation \cite{Winkler_text}. The cyclotron energy is the eigenvalue of the symmetric part of $\scr{H}_{n=0}$. A straightforward calculation yields $\hbar \wc= (e\hbar B/c)\sqrt{\det \alpha (\alpha^{-1})_{ii}}$, where $\alpha$ is the inverse mass tensor given by
\begin{equation}
	\alpha_{ij}=\sum_{n\neq 0}\frac{t_{ni} t_{nj}^* + t_{nj}t_{ni}^* + u_{ni}u_{nj}^* + u_{nj}u_{ni}^*}{E_0 -E_n}.
	\label{eq1}
\end{equation}
We have extended the notations of Cohen-Blount as $\bm{t}_n=\bm{v}_{0n}^{\up \up}$ and $\bm{u}_n=\bm{v}_{0n}^{\up\down}$, where $\bm{v}_{ij}^{\zeta \eta}$ is the interband matrix element of the velocity operator between $i$-th band with spin $\zeta$ and $j$-th band with spin $\eta$. $E_{\rm Z}$ is given as the eigenvalue of the antisymmetric part of $\scr{H}_{n=0}$. For $\bm{B}\parallel i$, it is obtained as $E_{\rm Z} = \pm (\tilde{g}/2)\mu_{\rm B} B_i$, where $\tilde{g} =2m\sqrt{G_{ii}}$,
\begin{equation}
	G_{ii}=4\left|\left(
	\sum_{n\neq 0} \frac{\bt_n \times \bu_n}{E_0 - E_n}
	\right)_i\right|^2
	-\left(
	\sum_{n\neq 0} \frac{\bt_n \times \bt_n^* + \bu_n \times \bu_n^*}{E_0 - E_n}
	\right)_i^2,
	\label{eq4}
\end{equation}
and $\mu_{\rm B}=e\hbar /2mc$. Then, $M$ for $B\parallel z$ is
\begin{equation}
	M = \frac{\D E_{\rm Z}}{\hbar \wc}
	=\sqrt{\frac{G_{zz}}{\alpha_{xx}\alpha_{yy}-\alpha_{xy}^2}}.
	\label{eq3}
\end{equation}
The obtained results \eqref{eq1}-\eqref{eq3} are all gauge independent. It should be stressed that the denominator $E_0-E_n$ of $\alpha$ and $G$ are approximately canceled each other in $M$. Therefore, $M$ is very sensitive to the symmetric properties of the interband matrix elements and insensitive to the energy differences. 
If only two bands ($n=0, 1$) are taken into account, $G_{zz}=\alpha_{xx} \alpha_{yy} - \alpha_{xy}^2$, and then $M$ would be exactly unity, consistent with previous results \cite{Cohen1960,Wolff1964,Fuseya2015}. 
If one take into account more than two bands, one can see that the form of Eqs. \eqref{eq1}-\eqref{eq3} already tells us that $M$ can be larger than unity. 

	Various theories of $\tilde{g}$ for multiband systems have been studied so far. The simple formula available at the present moment is valid only in the semiclassical limit \cite{Mikitik2002}, whereas the formulae for Bloch bands based on the quantum treatment are too complex to compute $\tilde{g}$ for various systems \cite{Blount1962,Roth1962,Yafet1963}. The present quantum formulae for $\hbar \wc$, $\tilde{g}$, and $M$ are general, rigorous within $\kp$ theory, and yet easy to handle. It is the advantage of these formulae that these values can be automatically obtained from the interband matrix elements and the energy differences, which can be directly computed by the band calculations as shown later.
	


	Here we give a concrete example of the above general theory. We adopt obtained formulae to the $T$-point holes in bismuth by taking into account the symmetry at the $T$-point. But it must be noted that the following arguments are also valid for group V semimetals (Sb, As), IV-VI narrow gap semiconductors (PbTe, PbSe, SnTe, etc.), and the topological insulator Bi$_2$Se$_3$, since the $k$-points where their carrier locate have the same symmetry as the $T$-point of bismuth \cite{Dimmock1964,HZhang2009}. The hole band at the $T$-point has the symmetry of $T_{45}^-$ \cite{Mase1958,Golin1968}. The symmetries of the other bands at the $T$ points are shown in Fig. \ref{Fig4} (e) and Fig. 1 in SM. (The group theoretical notation is that of Ref. \cite{Golin1968}.) From the selection rules, the matrix elements between the $T_{45}^-$ band and the others are finite only for
$\bm{t}_n^{(6)}=\langle T_{45}^- (1) | \bm{v} | T_6^+ (n) \rangle = (-a_n, i a_n, 0)$,
$\bm{u}_n^{(6)}=\langle T_{45}^- (1)  |\bm{v} | C T_6^+ (n) \rangle = (-a_n, -i a_n, 0)$,
and
$\bm{u}_n^{(45)}=\langle T_{45}^- (1) |\bm{v} | CT_{45}^+ (n) \rangle = (0, 0, b_n)$,
where $C$ is the product of space inversion and time reversal operators and $a_n$, $b_n$ are complex numbers \cite{Golin1968}. The $x$, $y$, and $z$-directions are taken along the binary, bisectrix, and trigonal axes, respectively. Since $\tilde{g}$ is given by the outer products of $\bt_n$ and $\bu_n$ [Eq. \eqref{eq4}], it is clear from $\bm{t}_n^{(6)}$ and $\bm{u}_n^{(45), (6)}$ that  $G_{zz}$ is the only non-vanishing term. Therefore, the angular dependence of $\D E_{\rm Z}$ is simply $\D E_{\rm Z}(\theta)=2m\mu_{\rm B}B|\cos \theta|\sqrt{G_{zz}}$, where $\theta$ quantifies the tilt angle of the magnetic field off the trigonal axis. This leads to $\tilde{g}=0$ for $B\perp$ trigonal axis, providing a solution to the first experimental puzzle.

	The ratio $M$ for the $T$-point holes is obtained as
\begin{equation}
	M= \sqrt{
	\left|
	\sum_{n\neq 0} \frac{a_n^2}{E_0 -E_n}
	\right|^2
	\Bigg/\displaystyle
	\left(
	\sum_{n\neq 0} \frac{|a_n|^2}{E_0 - E_n}
	\right)^2
	}.
	\label{eq8}
\end{equation}
This form makes it easy to understand how $M$ can exceed unity. For example, in the case with three bands $n=0, 1, 2$, whose energies are $E_2 < E_0 < E_1$, it is easy to show that the ratio is always greater than one \emph{no matter how large the energy differences are}  [Eq. (32) in SM]. Accordingly, one expects that the interband contributions from the lower (higher) energy bands increase (decrease) the ratio, providing a possible solution to the second experimental puzzle.

	In order to know the definite value of $M$, we need to evaluate the interband matrix elements $\bm{t}_n$, $\bm{u}_n$ and the energy differences $E_0 - E_n$ based on the band calculations. Here we evaluate them based on the multiband $\kp$ Hamiltonian (eight-band model) derived from the well-known tight-binding band calculation of Bi by Liu and Allen, which is in quantitative agreement with experiments \cite{Liu1995}. Then,  using the obtained formulae for $\alpha_{ij}$, $\tilde{g}$, and $M$, they can be automatically obtained as $\alpha_{xx}=\alpha_{yy}=14.1$, $\alpha_{xy}=0$, $\tilde{g}=58.7$, and $M = 2.08$ for $B\parallel {\rm trigonal}$, which agrees well with the experimental value of $2.12$. The only band that can increase $M$ is $T_6^+ (2)$ (Fig. 1 in SM). Therefore, and surprisingly, a band 1eV far from the band in which carriers resides can enhance the magnitude of $M$ by a factor of two. This large interband effect provides a quantitative solution to the second puzzle.


\begin{figure}
\begin{center}
\includegraphics[width=7cm]{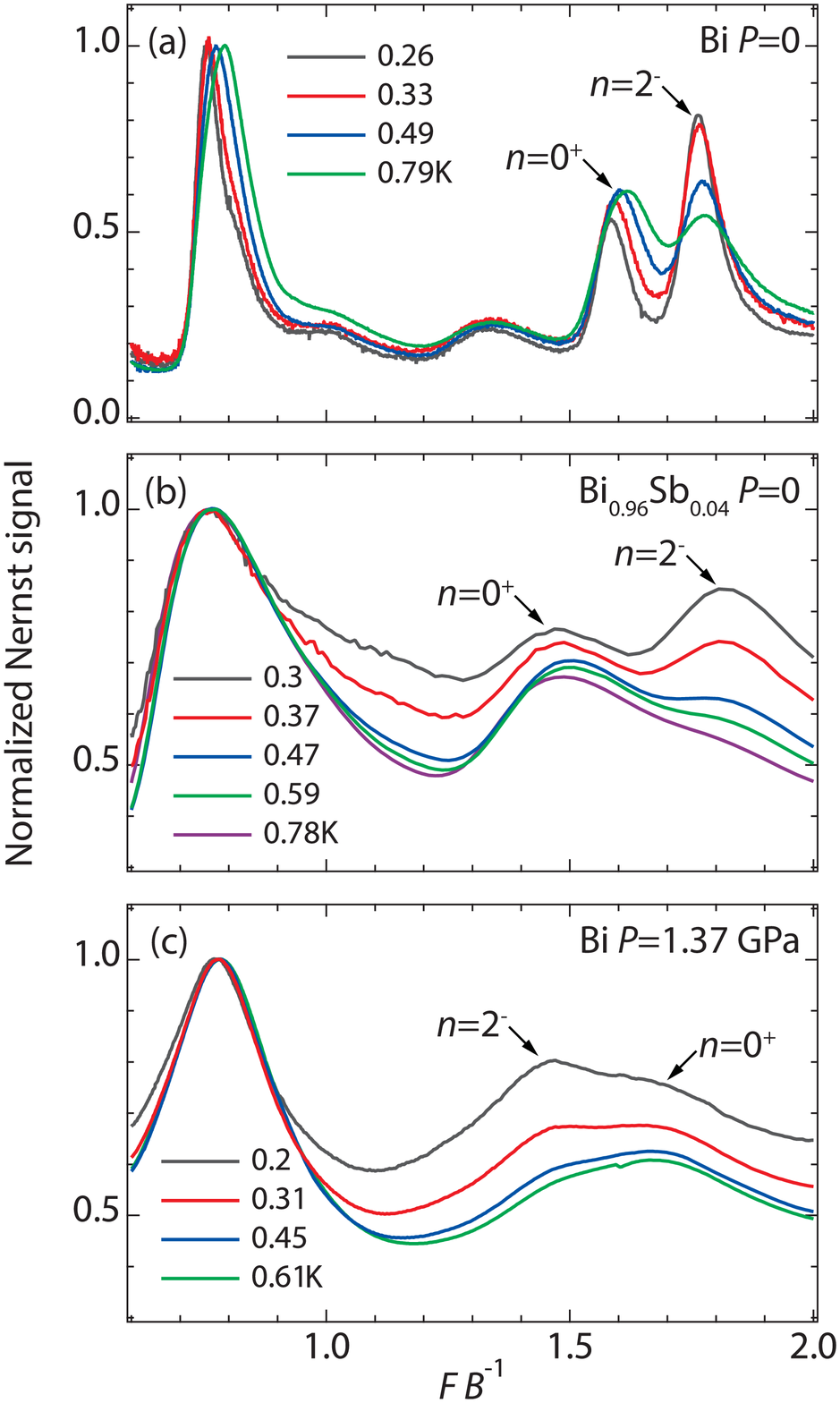}
\end{center}
\caption{\label{Fig3}Nernst signal as a function of $FB^{-1}$ at different temperatures for (a) pure Bi at $P=0$, (b) Bi with 4\% Sb substitution at $P=0$, and (c) pure Bi at $P=1.37$ GPa. The magnetic field is along the trigonal axis and $F$ is the frequency of the quantum oscillation. Peaks are identified by their Landau sub-level indexes.}
\end{figure}

\begin{figure}
\begin{center}
\includegraphics[width=8cm]{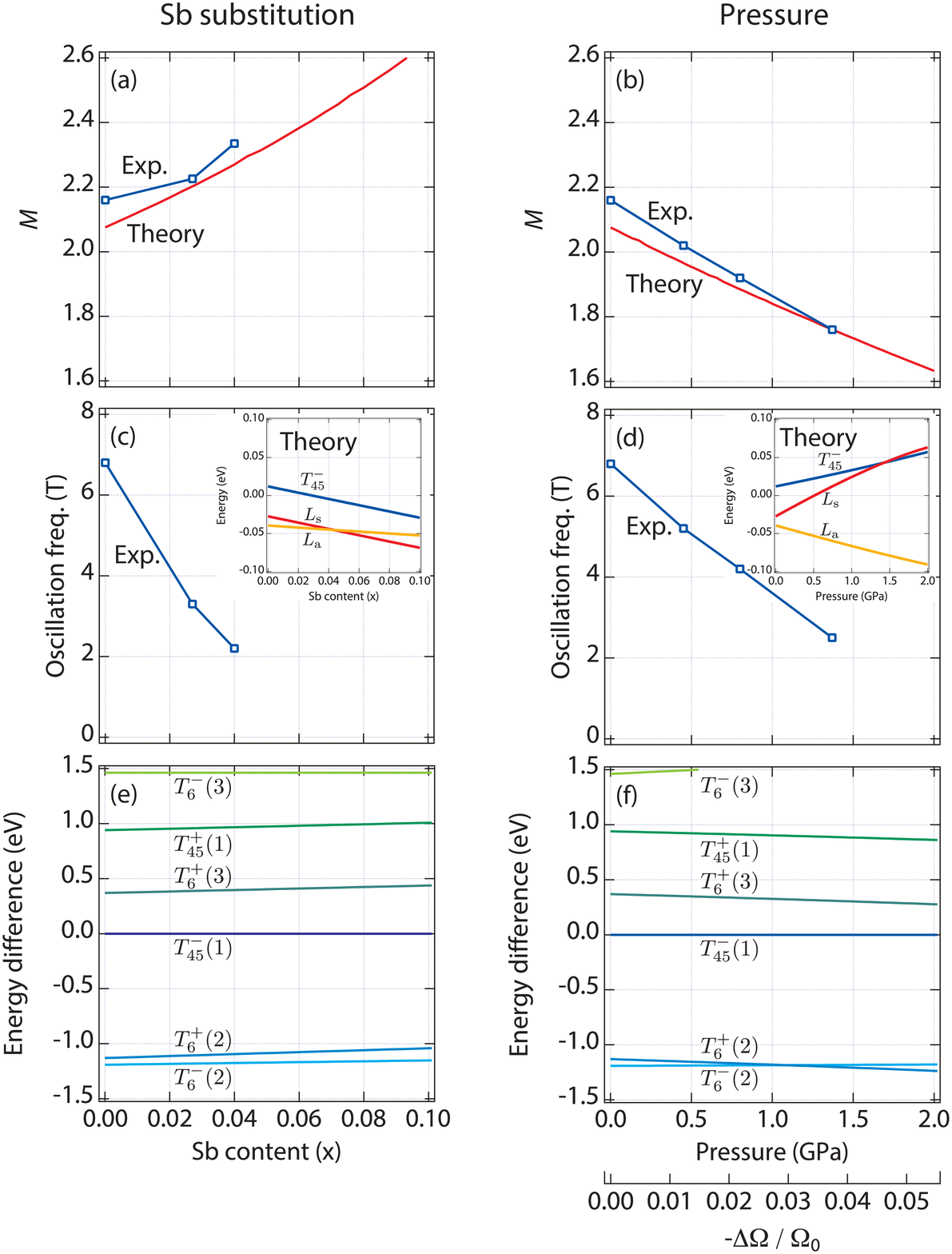}
\end{center}
\caption{\label{Fig4}The ratio $M$, the quantum oscillation frequency, and the energy differences from the hole band ($T_{45}^-$) as a function of Sb content (left column) and pressure (right column). Experimental and theoretical results are shown by symbols and solid lines, respectively. For pressure, the theoretical results are obtained with respect to the compression in volume, $-\Delta \Omega /\Omega_0$. ($-\Delta \Omega /\Omega_0 = 0.041$ corresponds to 1.5 GPa \cite{Giardini1965}).
Insets of (c) and (d) are the energy shift of the $T$-point hole band, the conduction and valence band at the $L$-point, taken the origin of the energy as the position of the Fermi energy of pure Bi at zero pressure.}
\end{figure}

	An additional cross-check is provided by the evolution of $M$ with alloying \cite{Lenoir1996} or by applying pressure. One may naively expect that substituting bismuth with antimony can be assimilated to chemical pressure and therefore the results should be identical to applying physical pressure. But there are two essential differences. First, the substitution changes the strength of the SOI, while the pressure does not. Second, the lattice structure becomes more rhombohedral by substitution \cite{Dresselhaus1971}, while it tends to approach cubic by the pressure \cite{Aoki1982}.  Figure \ref{Fig3} shows the Nernst signal of Bi as a function of $F B^{-1}$ for pure Bi at $P=0$, Bi$_{0.96}$Sb$_{0.04}$ alloy at $P=0$, and pure Bi at $P=1.37$ GPa. (See SM for details on experiments.) We can distinguish the peaks of $n=0^+$ and $2^-$ from their different temperature dependence. In the case of Sb substitution, the $0^+$ and $2^-$ peaks shift, but the order is unchanged. Applying  pressure, on the other hand, inverts the order. This difference can be seen by plotting $M$ as a function of Sb content and pressure shown in Figs. \ref{Fig4} (a) and (b), respectively. $M$ increases with substitution but decreases with pressure. Note that in both cases as shown in Figs. \ref{Fig4} (c) and (d) the frequency of quantum oscillations decreases as it was found before \cite{Brandt1971,Banerjee2008}. This implies that the Fermi surface shrinks because the overlap between the conduction band at the $L$-point and the valence band at the $T$-point is reduced.

	These experimental results can be naturally interpreted by the present theory in terms of the relevance of the interband contributions. Figures \ref{Fig4} (e) and (f) show the theoretically obtained energy differences from the hole band ($T_{45}^-$) with respect to the Sb content and the compression in volume, $-\Delta \Omega /\Omega_0$, respectively. ($-\Delta \Omega /\Omega_0 = 0.041$ corresponds to 1.5 GPa \cite{Giardini1965}.) In the case of the substitution, both the higher energy $T_6^+ (2)$, $T_{45}^+ (1)$ and the lower energy $T_6^+ (1)$ go up, so that the contribution from the lower energy band increases resulting in the enhancement of $M$ shown by lines in Fig. \ref{Fig4} (a). In the case of pressure, on the other hand, both the higher and lower bands move downward, so that the contribution from the higher energy band increases resulting in the decrease of $M$ shown in Fig. \ref{Fig4} (b). Even though the energy shifts are very small, they make a sizable change in $M$. In both cases, the overlap between the $L$-conduction and $T$-valence bands decreases as shown in the insets of Figs. \ref{Fig4} (c) and (d). As seen in the figure, there is a satisfactory agreement between theory and experiment. For Sb substitution, we adopt a simple virtual crystal approximation \cite{Teo2008}. As for pressure, we assume that the overlap integrals scale with $d^{-2}$, where $d$ is the bond length \cite{Liu1995}. (See Sec. V and VI in SM for details.) 

	In summary, we studied both experimentally and theoretically the ratio of the Zeeman splitting to the cyclotron energy $M$, which characterizes the effect of crystalline SOI. A general theory on $M$ was newly derived based on the multiband $\kp$ theory. By this multiband approach, we succeeded in solving the longstanding mystery for hole-like carriers at the $T$-point of bismuth, the large and anisotropic $M$, which have not been explained by the previous two-band approach. Our results strongly suggest the surprisingly large impact of interband effect of the crystalline SOI. Moreover, we gave new experimental results on $M$ by substituting antimony and applying pressure, both were quantitatively explained by the present multiband $\kp$ theory. The quantitative agreement with experiment is another success for the present multiband $\kp$ theory. Beyond the specific case of bismuth, this general scheme can be applied to all materials where SOI plays a relevant role, such as group V semimetals (Sb, Ab), IV-VI narrow gap semiconductors (PbTe, PbSe, SnTe, etc.), and the topological insulator (Bi$_2$Se$_3$) in which $M\sim2$ has been reported\cite{Kohler1975,Fauque2013, Orlita2015}.

\bibliography{Bismuth.bib}
\end{document}